\documentclass[conference,letterpaper]{IEEEtran}

\usepackage[utf8]{inputenc} 
\usepackage[T1]{fontenc}
\usepackage{url}
\usepackage{ifthen}
\usepackage{cite}
\usepackage{graphicx}

\usepackage[cmex10]{amsmath}
\usepackage{amsfonts,amssymb} 

\newtheorem{lemma}{\textbf{Lemma}}

\begin{document}
\title{An Orthogonality Principle for Select-Maximum Estimation of Exponential Variables} 



\author{\IEEEauthorblockN{Uri Erez}
\IEEEauthorblockA{\textit{Dept.\ of Electrical Engineering-Systems} \\
\textit{Tel Aviv University}\\
Tel Aviv, Israel \\
urierez@tauex.tau.ac.il}
\and
\IEEEauthorblockN{Jan \O stergaard}
\IEEEauthorblockA{\textit{Dept.\ Electronic Systems} \\
\textit{Aalborg University}\\
Aalborg, Denmark \\
jo@es.aau.dk}
\and
\IEEEauthorblockN{Ram Zamir}
\IEEEauthorblockA{\textit{Dept.\ of Electrical Engineering-Systems} \\
\textit{Tel Aviv University}\\
Tel Aviv, Israel \\
ramzamir@tauex.tau.ac.il}}


\maketitle

\begin{abstract}
Motivated by multiple-description source coding with feedback, it was recently proposed to encode the one-sided exponential source $X$ via $K$ parallel channels, $Y_1,\dotsc, Y_K,$ such that the 
error signals $X-Y_i, i=1,\dotsc, K,$ are one-sided exponential and mutually independent given $X$. Moreover, it was shown that the optimal estimator $\hat{Y}$ of the source $X$ with respect to the one-sided error criterion, is simply given by the maximum of the outputs, i.e., $\hat{Y} = \max \{ Y_1,\dotsc, Y_K\}$. In this paper, we show that the distribution of the resulting estimation error $X-\hat{Y}$, is equivalent to that of the optimum noise in the backward test-channel of the one-sided exponential source, i.e., it is one-sided exponentially distributed and statistically independent of the joint output $Y_1,\dotsc, Y_K$. 
\end{abstract}


\section{Introduction}
\begin{figure}[th]
\begin{center}
\includegraphics[width=4cm]{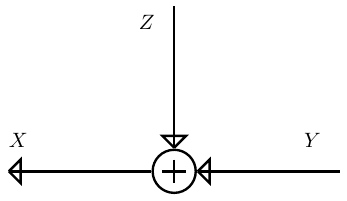}
\caption{Backward additive test channel. The reconstruction $Y$ and the channel noise $Z$ are mutually independent.}
\label{fig:test-channel}
\end{center}
\end{figure}

The rate-distortion function $R(D)$ for a source $X$ under the distortion measure $d(x,y)$ is defined as the infimum of the mutual information $I(X;Y)$ over the set $\{f_{Y|X}(y|x)\}$
of conditional probability density functions, which satisfy a given distortion constraint $\mathbb{E}[d(X,Y)]\leq D$ \cite{gallager:1968,berger:1971}. The optimum distribution, say $f^*_{Y|X}(y|x)$, can be expressed as $f^*_{Y|X}(y|x) =
f^*_{X|Y}(x|y)f_Y(y)/f_X(x)$, where 
the conditional density $f^*_{X|Y}(x|y)$ is often described via a so-called \emph{backward} test channel \cite{gallager:1968,berger:1971}, where the reconstruction $Y$ acts as the \emph{input} and the source $X$ is the \emph{output} of the test channel, see Fig.~\ref{fig:test-channel}.

For several sources and distortion measures, it is known that the backward channel is an additive channel in the sense that the channel noise $Z$ is statistically independent of $Y$, and the source is simply given as the sum $X=Z+Y$. Moreover, in some cases $X$ and $Z$ belongs to the same family of distributions. For example, in the case of a Gaussian source $X$ and under the mean squared error distortion measure, the optimum channel noise $Z$ in the backward test channel is Gaussian and independent of the reconstruction $Y$ \cite{gallager:1968,berger:1971}. A similar relationship can be observed for the binary source under the hamming distortion measure~\cite{berger:1971} and for the one-sided exponential source under the one-sided error distortion measure~\cite{Ver96}. 

If we have access to $K\geq 1$ different encodings of a Gaussian source, for example  $Y_i = X + N_i, i=1,\dotsc, K$, where $X$ and $N_1,\dotsc, N_K$ are jointly Gaussian, then the estimation error $X - g(Y_1,\dotsc, Y_K)$ due to optimally estimating $X$ from $Y_1,\dotsc, Y_K$, in a minimum mean-squared error (MSE) sense, is also Gaussian. Moreover, this estimation error is independent of the outputs $Y_1,\dotsc, Y_K$. 

In this paper, we are interested in the case, where we have access to $K \geq 1 $ outputs (reconstructions) $Y_i$, $i=1,\dotsc, K,$ of a one-sided exponential source $X$. 
Each reconstruction is by itself optimal under the one-sided error distortion measure and therefore satisfies the backward  channel relationship $X = Z_i + Y_i$,  where $Z_i$, $i=1,\dotsc,K$ are one-sided exponentially distributed. We assume that the reconstructions $Y_1,\dotsc, Y_K,$ are \emph{independent} encodings, which is a concept introduced in \cite{erez:2020,ostergaard:2020}, and which implies that the outputs $Y_i, i=1,\dotsc, K,$ are mutually independent given $X$. Thus, the following Markov chains apply: $Y_i - X - Y_j, \forall i,j$. 
It was shown in \cite{erez:2020} that the optimum estimator $g(Y_1,\dotsc, Y_K)$ of $X$ given $Y_1,\dotsc, Y_K$, under the one-sided error distortion measure is the maximum of $Y_1,\dotsc, Y_K$, i.e., 
$$g(Y_1,\dotsc, Y_K) = \max_{i} Y_i.$$ 

We show in this work, that the estimation error $X - g(Y_1,\dotsc, Y_K)$ has a backward test channel formulation. Specifically, the estimation error $X - g(Y_1,\dotsc, Y_K)$ is independent of the output vector $(Y_1, \dotsc, Y_K)$ and is one-sided exponentially distributed. 

Our key motivation for considering the statistical properties of the estimation error is related to the problem of multiple descriptions with feedback~\cite{ostergaard:2020}. Note that the $K$ independent encodings can be seen as a multiple-description code in the case of no excess marginal rate~\cite{Zhang1995MultipleDS}. Moreover, assume that the encodings (descriptions) are transmitted individually over a lossy network, and that the decoder is able to inform the encoder which descriptions  were received. For sake of refinement in the next round of communication, the encoder will  estimate the source using only the particular subset of  descriptions that were received at the decoder. As is shown in this paper, the estimation error will be one-sided exponential distributed and will be independent of the descriptions that were not lost. One may therefore encode the estimation error into $K$ new independent encodings (multiple descriptions) and thereby further refine the estimate of the source at the decoder. 

The sum-rate of the $K$ "independent" descriptions generally exceeds the rate-distortion function, which is obtained with a single description yielding the same distortion as that obtained with the $K$ descriptions. However, as was shown in \cite{ostergaard:2020}, if the sum-rate is very small, the rate loss is also very small. Specifically, assume the source is encoded over $M$ rounds with $K$ descriptions in each round. Asymptotically as the number of rounds tends to infinity and the sum-rate per round vanishes, then the accumulated sum-rate over all rounds is finite and non-zero, and
 it becomes rate-distortion optimal to use $K$ independent descriptions in each round \cite{ostergaard:2020}.

It is interesting to note that the estimation error, when decoding using any subset of the $K$ descriptions will become one-sided exponentially distributed. 
However, the average estimation error due decoding using a \emph{random} subset of the descriptions will not be one-sided exponentially distributed. We provide a closed-form expression for the distribution of the average estimation error in the case of using a random number of encodings for estimating the source. 

\subsection{Notation}
We use upper case letters for random variables, and lower case letters for their realizations. For random variables $X,Y$, we use the notations $f_{Y|X}, F_{Y|X}, CF_{Y|X}$ for the conditional probability density function (PDF), conditional cumulative distribution function (CDF), and conditional complimentary CDF of $Y$ given $X$, respectively. 

\section{Background}
The one-sided exponential source with parameter $\lambda>0$ is defined as \cite{Ver96}:
\begin{equation}\label{eq:source}
f_X(x) = \lambda e^{-\lambda x}, \quad x\geq 0,
\end{equation}
where $\mathbb{E}{[X]} = 1/\lambda$. 

Let $y$ denote the reproduction of the coder, and let the one-sided  error criterion be given by:
\begin{equation}\label{eq:distortion}
  d(x,y) = \begin{cases}
  x-y, & \text{if}\  x\geq y \geq 0, \\
  \infty, &  \text{if}\ x<y. 
  \end{cases}
\end{equation}

The rate-distortion function (RDF) for the exponential source with one-sided error criterion is given by \cite{Ver96}:
\begin{equation}
R(D) = \begin{cases}
-\log(\lambda D), & 0\leq D\leq \frac{1}{\lambda}, \\
0, & D> \frac{1}{\lambda}.
\end{cases}
\end{equation}

Let $X=Z+Y$ denote the backward test channel whose optimal conditional output distribution is given by \cite{Ver96, si:2014}:
\begin{equation}\label{eq:cond}
f_{X|Y}(x|y) = \frac{1}{D} e^{-\frac{(x-y)}{D}}, \quad x\geq y\geq 0.
\end{equation}
The channel noise, $Z$, is one-sided exponential distributed with parameter $1/D$. The distribution of $Z$ is easily obtained from \eqref{eq:cond} by inserting $Z=X-Y$. 

The output distribution for $Y$ is a mixture distribution with an atom at zero and is given by \cite{erez:2020}:
\begin{equation} \label{eq:fy}
f_Y(y) = 
\begin{cases}
\lambda D \delta(y) + (1-\lambda D) \lambda e^{-\lambda y} ,& \quad 0\leq y\leq x, \\
0, & \text{otherwise}. 
\end{cases}
\end{equation}

If we take the channel from $X$ to $Y$ to be the RDF-achieving forward test channel, then: 
\begin{align}\label{eq:FYX}
F_{Y|X}(y|x)&= e^{\delta(y-x)}, 
\end{align}
where $\delta = \frac{1}{D} - \lambda$, and $0\leq y\leq x$. 

\section{Distribution of the Estimation Error}
Let $X$ be one-sided exponentially distributed, and consider $K$ parallel test-channels, $Y_1,\dotsc, Y_K$, where the errors $Z_i = X - Y_i, i=1,\dotsc, K,$ are mutually independent given $X$. 
We will refer to the set $(Y_1,\dotsc, Y_K)$ as parallel channels or independent encodings of the source $X$ \cite{erez:2020}. 

It was shown in \cite{erez:2020}, that the select-max estimator, i.e., the estimator that simply selects the maximum  of the $K$ channels as the estimate of $X$, 
$$\hat{Y} = \max \{ Y_1,\dotsc, Y_K\},$$
is in fact an optimal estimator under the one-sided error distortion. 

The following  lemmas show that the estimation error is one-sided exponentially distributed and that it is independent of the outputs $(Y_1,\dotsc, Y_K)$. 

\vspace{1mm}

\begin{lemma}[\textbf{Exponential estimation error}]
\emph{
Let $Y_1,\dotsc, Y_K$, be $K\geq 1$ independent encodings of the one-sided exponential source $X$ with parameter $\lambda>0$. Moreover, 
let $\delta = \frac{1}{D} - \lambda, 0\leq D\leq 1/\lambda$. 
Then, the estimation error $\tilde{Z}=X - \hat{Y}$, due to estimating the source $X$ by the select-max estimator $\hat{Y}$, is one-sided exponentially distributed with parameter $\lambda' = \lambda+ K\delta$, i.e.:
\begin{equation}
f(\tilde{z}) = (\lambda+K\delta) e^{-(\lambda+K\delta)\tilde{z}},\quad  \tilde{z}\geq 0.
\end{equation}
}
\end{lemma}

\begin{IEEEproof}[\textbf{Proof}]
The sequence of outputs  $Y_1, \dotsc, Y_K,$ are absolute continuous and conditionally independent given $X$. The conditional CDF of the select-max estimator from the $K$ channels is therefore given by \cite{erez:2020}:
\begin{align}
F_{\hat{Y}|X}(\hat{y}|x) &= F_{Y|X}(y|x)^K,
\end{align}
where $F_{Y|X}(y|x)$ is the conditional CDF of each channel, respectively.  We have here omitted the channel index $i$, since $Y_i, \forall i,$ are identically distributed. 
The conditional PDF $f_{\hat{Y}|X}(\hat{y}|x)$ is given by:
\begin{equation}
f_{\hat{Y}|X}(\hat{y}|x) = K F_{Y|X}(y|x)^{K-1}f_{Y|X}(y|x).
\end{equation}
Using \eqref{eq:FYX} it now follows that:
\begin{align}\label{eq:CDF_YX}
F_{\hat{Y}|X}(\hat{y}|x) = e^{ K \delta (y-x) },  0\leq y \leq x.
\end{align}

Since the estimation error is given by $\tilde{Z} = X - \hat{Y}$, the probability of the event 
$\tilde{Z}\geq \xi$ is equal to that of the event $\hat{Y}\leq \xi$, for some $\xi \in \mathbb{R}_+$. So the complementary CDF of $\tilde{Z}$ 
and the regular CDF of $\hat{Y}$ are the same (up to a shift by $X$). 
It follows that the conditional complimentary CDF $CF_{\tilde{Z}|X}(\tilde{z}|x)$ is given by  
\begin{align} \label{eq:CF}
CF_{\tilde{Z}|X}(\tilde{z}|x)& = F_{\hat{Y}|X}(\hat{y}=x-\tilde{z}|x)\\
& = e^{-K \delta \tilde{z}},\quad  0\leq \tilde{z} \leq x,
\end{align}
and the unconditional complimentary CDF $CF_{\tilde{Z}}$ is given by:
\begin{align}\label{eq:uccdf0}
CF_{\tilde{Z}}(\tilde{z}) &= \int_{x=\hat{y}}^{\infty} CF_{\tilde{Z}|X}(\tilde{z}|x) f(x) dx \\
&=\int_{x=\hat{y}}^{\infty} e^{K\delta(\hat{y}-x)} \lambda e^{-\lambda x} dx \\ \label{eq:uccdf}
&= e^{ - (\lambda + K \delta) \tilde{z}}.
\end{align}
The unconditional CDF of the estimation error when using the select-max estimator on the $K$ channel outputs can now easily be obtained from the unconditional complimentary CDF in \eqref{eq:uccdf}, that is:
\begin{align}
F_{\tilde{Z}}(\tilde{z}) = 1 - e^{ - (\lambda + K \delta) \tilde{z}},\quad  \tilde{z} \geq 0, 
\end{align}
which implies that the error is exponentially distributed:
\begin{align}\label{eq:distr}
f_{\tilde{Z}}(\tilde{z}) = \frac{d}{d \tilde{z}} F_{\tilde{Z}}(\tilde{z}) = (\lambda+K\delta) e^{-(\lambda+K\delta)\tilde{z}}, \tilde{z}\geq 0,
\end{align}
This proves the lemma. 
\end{IEEEproof}

\vspace{1mm}

\textbf{Remark:}
\emph{The  distortion $D_K$ of the estimator $\hat{Y}$ in Lemma 1 can be written as:}
\begin{align}
\frac{1}{D_K} = \lambda + K \bigg(\frac{1}{D} - \lambda\bigg),
\end{align}
which resembles the expression for the MMSE formula for $K$ measurements. 

\vspace{1mm}

\begin{lemma}[\textbf{Sufficient statistics}]
\emph{Let $X$ be a one-sided exponential source, and let $X \to Y_1, \dotsc, X \to Y_K$ be $K$ parallel (RDF achieving) test channels. 
Moreover, let  $\hat{Y} = \max \{Y_1,\dotsc , Y_K\}$. 
Then,
\begin{align}
X -  \hat{Y} - (Y_1, \dotsc, Y_K) 
\end{align}
form a Markov chain, i.e., $\hat{Y}$ is a sufficient statistic for $X$ from $(Y_1, \dotsc, Y_K)$.}
\end{lemma}


\begin{IEEEproof}[\textbf{Proof}]
In order to prove the lemma, we will  derive an explicit expression for the conditional distribution
of $X$ given $(Y_1, \dotsc, Y_K)$, and observe that it only depends on the maximum value $\hat{Y}$ of the vector $(Y_1, \dotsc, Y_K)$. Then we use this result to further show that the conditional distribution of $X$ given $\hat{Y}$ and $(Y_1, \dotsc, Y_K)$ does not depend upon $(Y_1, \dotsc, Y_K)$. 

%

Let $y>0$ be any positive number greater than $\tilde{y}_2, \dotsc, \tilde{y}_K$, i.e., 
$y > \tilde{y}_2, \dotsc, y > \tilde{y}_K$, where $\tilde{y}_i > 0, i=2, \dotsc, K$.  
The conditional density $f(x | Y_1=y, Y_2 < \tilde{y}_2,\dotsc, Y_K < \tilde{y}_K)$ of the source $X$ given the event $\{Y_1=y, Y_2 < \tilde{y}_2,\dotsc, Y_K < \tilde{y}_K\}$ on the $K$ outputs is 
 nonzero only for $x \geq y$, since the test channels' outputs are smaller than $X$ with probability 1. 
%
%
By the complete probability formula (Bayes law) this conditional density is equal to:
\begin{align} 
f(x | &Y_1=y, Y_2 < \tilde{y}_2,\dotsc, Y_K < \tilde{y}_K) \\ \notag
&= f(x) f(y|x) F(\tilde{y}_2|x) \cdots F(\tilde{y}_K|x) \\   \label{eq:csd}
&\times 
\bigg(
\int_{x=y}^{\infty}
f(x) f(y|x) F(\tilde{y}_2|x) \cdots F(\tilde{y}_K|x) \bigg)^{-1},
\end{align}
where we used the fact that all the channels are identical and also conditional independent given the source $X$. 
Inserting the conditional commulative distribution $F(y|x)$ of each channel, which is given by \eqref{eq:FYX}, makes it possible to rewrite the numerator (i.e., the terms outside the big brackets) of \eqref{eq:csd} as follows:
\begin{align}
 f(x) f(y|x) & F(\tilde{y}_2|x) \cdots F(\tilde{y}_K|x) \\
  &= f(x) f(y|x) e^{-(x-\tilde{y}_2) \delta} \cdots e^{-(x-\tilde{y}_K) \delta}  \\
  &= f(x) f(y|x) e^{-(K-1) x \delta} e^{ \tilde{y}_2 \delta} \cdots e^{\tilde{y}_K \delta},
\end{align}
where the factor $e^{\tilde{y}_2 \delta} \cdots e^{\tilde{y}_K \delta}$  is independent of $x$. We can rewrite the denominator (the terms inside the big brackets) of \eqref{eq:csd} in a similar manner, where the factor $e^{\tilde{y}_2 \delta} \cdots e^{\tilde{y}_K \delta}$ also appears and goes outside the integral to be cancelled by the equivalent factor in the numerator.  
Hence the conditional density is simply given as:
\begin{align} \label{eq:fxy}
& f(x | Y_1=y, Y_2 < \tilde{y}_2, \dotsc, Y_K < \tilde{y}_K) \\ \label{eq:fxy1}
&= f(x) f(y|x) e^{-(K-1) x \delta} \bigg( \int f(x) f(y|x) e^{-(K-1) x \delta} dx\bigg)^{-1},
\end{align}
which is indeed independent of $\tilde{y}_2,\dotsc, \tilde{y}_K$ as desired.

If we now let the maximal output $Y_{i^*} = \hat{Y}$, where $i^*=\arg\max_i \{Y_i\}$ be equal to $y$, i.e., $Y_{i^*} =y$, and the remaining $K-1$ outputs, $Y_1, \dotsc,  Y_{i^*-1}, Y_{i^*+1}, \dotsc, Y_K$ be smaller than $\tilde{y}_2, \dotsc, \tilde{y}_K$, respectively, then we can form the following $K$ non-intersecting events:
\begin{align} \label{eq:y1}
\{ Y_1 &= y, Y_2 < \tilde{y}_2, \dotsc, Y_K < \tilde{y}_K \} \\
\{ Y_2 &= y, Y_1 < \tilde{y}_2,  Y_3 < \tilde{y}_3, \dotsc,  Y_K < \tilde{y}_K \} \\ \notag
 \qquad  \vdots \\ 
\{ Y_K &= y, Y_1 < \tilde{y}_2, \dotsc, Y_{K-1} <  \tilde{y}_K  \}.
\end{align}
Consider the first event given in \eqref{eq:y1}, which is also the event occuring in \eqref{eq:fxy} -- \eqref{eq:fxy1}. This event depends only on $x, y$ and $K$, but is independent of $(\tilde{y}_2, \dotsc, \tilde{y}_K)$.  
Due to symmetry, the probabilities of the events are equal, and they are each individually independent of $(\tilde{y}_2, \dotsc, \tilde{y}_K)$, which implies that so is the union of the events.  
This proves the lemma for the case where $Y_i>0, \forall i$. 

We have yet to consider the case where $Y_i = 0$ for some $i$. If $\hat{Y} = 0$, then clearly all channel outputs are zero, and it easily follows that $X- \hat{Y} - (Y_1, \dotsc, Y_K)$ is satisfified. Let $\hat{Y}\neq 0$ but $Y_i=0$ for some $i$, say $i=1$. Then, in this case we have:
\begin{align*}\notag
&f_{Y|X}(y_1=0|x) = f_{X|Y}(x|y_1=0) f_{Y}(y_1=0) f_X(x)^{-1} \\ 
&=  \frac{1}{D}e^{-x/D}   \lambda D f_X(x)^{-1} =  \lambda e^{-x/D}   f_X(x)^{-1}\\
&=  \lambda e^{-x (\delta + \lambda)}   f_X(x)^{-1}  =  e^{-x \delta } =  F_{Y|X}(y=0|x),
\end{align*}
where we used $D^{-1}= \lambda+\delta$. The lemma is now proved since for $y=0, F_{Y|X}(y|x) = f_{Y|X}(y|x)$, which means that \eqref{eq:fxy} -- \eqref{eq:fxy1} are unaffected if one changes the event from $Y_i < \tilde{y}_i$, where $\tilde{y}_i>0$ into $Y_i = \tilde{y}_i = 0$.  
\end{IEEEproof}

\vspace{1mm}
\textbf{Remark:}
\emph{
It may be noticed that the conditional density $f(x|\hat{Y})$ simply describes the equivalent additive backward channel:
\begin{align}
 f(x | \hat{Y} = y) = f(x|\hat{y}) =f_{\tilde{Z}}(x - \hat{y}),
\end{align}
because $\tilde{Z}$ is exponential with a parameter $\lambda + K \delta$. Thus, if the source is exponential, then the conditional density $f(x|\hat{y})$ becomes 
$f_{\tilde{Z}}( x - \hat{y} )$, which is equivalent to the estimation error density computed at $\tilde{z} = x-\hat{y}$. 
}

\vspace{1mm}

\begin{lemma}[\textbf{Orthogonality principle}]
\emph{Let $X$ be a one-sided exponential source, and let $X \to Y_1, \dotsc, X \to Y_K$ be $K$ parallel (RDF achieving) test channels. Finally, let  $\hat{Y} = \max \{Y_1,\dotsc , Y_K\}$. 
Then, the backward channel $X = \hat{Y} + \tilde{Z}$ is additive, i.e., the estimation error $\tilde{Z}$ is independent of the estimator $\hat{Y}$. Moreover, $\tilde{Z}$ is independent of the joint output vector $(Y_1,\dotsc, Y_K)$. }
\end{lemma}

\vspace{1mm}

\begin{IEEEproof}[\textbf{Proof}]
That $\tilde{Z}=X-\hat{Y}$ is independent of $(Y_1,\dotsc, Y_K)$ follows immediately  due to $\hat{Y}$ being a sufficient statistics for $X$ from $(Y_1,\dotsc, Y_K)$, see Lemma 2. 

The first part of the lemma follows from the fact that the equivalent forward channel  $X \to \hat{Y}$ has the form: 
\begin{align}
F(\hat{y}|x) = F(y|x)^K = e^{-(x-y) K \delta},
\end{align}
which is the same form as $F(y|x)$, if we replace $\delta$ by $K \delta$. Thus, if the combination of an exponential source and the test channel $F(y|x)$ implies an additive exponential backward channel with distortion $(\lambda+\delta)^{-1}$, then the combination of an exponential source and the channel $F(y|x)^K$ implies an exponential additive backward channel with distortion $(\lambda+K \delta)^{-1}$. 
\end{IEEEproof}

\vspace{1mm}

\textbf{Remark:}
\emph{We note that the hypothetical rate $I(X; \hat{Y})$ is RD optimal w.r.t. the one-sided error distortion. This follows since the estimation error $X-\hat{Y}$ satisfies the "backward orthogonality" property, i.e., it is exponentially distributed and independent of $\hat{Y}$. Thus, we can write:
\begin{align*}
I(X;\hat{Y}) &= h(X) - h(X|\hat{Y}) = h(X) - h(X-\hat{Y}|\hat{Y})  \\
&= h(X) - h(X - \hat{Y})  = h(X) - h(\tilde{Z}) = R(D),
\end{align*}
where $\tilde{Z}$ is exponentially distributed. Moreover, since $X - \hat{Y} - (Y_1, \dotsc, Y_K)$ it follows that $I(X; Y_1,\dotsc, Y_K) = I(X; \hat{Y})$. Thus, $I(X; Y_1,\dotsc, Y_K)$ is also RD optimal w.r.t., the one-sided error distortion. Of course, the rate in independent encoding $I(X;Y_1) + \cdots + I(X;Y_K) = K I(X;Y_1)$ is  generally not RD-optimal, except at very small coding rates \cite{erez:2020}.}

\section{The case of an unknown number of encodings}
In the previous section, we assumed the availability of $K$ descriptions and formed the estimate $\hat{Y} = \max \{Y_1,\dotsc, Y_K\}$. 
If the $K$ descriptions are individually transmitted over a packet-switched network, where packets could occasionally be dropped, then the number of received descriptions, say $\ell$, becomes a random variable, where $\ell \in \{1,\dotsc, K\}$. 

Let us consider a packet erasure channel with packet-loss probability $\theta$, and where the packet losses are independently distributed. In this case, the probability that $\ell$ out of $K$ packets are received is given by:
\begin{equation}
\mathbb{P}(K=\ell) = \theta^{K-\ell}(1-\theta)^{\ell}.
\end{equation}
The conditional CDF of $\hat{Y}$ given $X$ and the event that $\ell$ descriptions are received is easily obtained from \eqref{eq:CDF_YX}:
\begin{align}
F_{\hat{Y}|X,\ell}(\hat{y}|x,\ell) = e^{ \ell \delta (y-x) },  0\leq y \leq x, 0\leq \ell \leq K.
\end{align}
By similar arguments leading to \eqref{eq:CF}, we obtain that:
\begin{align}
CF_{\tilde{Z}|X,\ell}(\tilde{z}|x,\ell)& = F_{\hat{Y}|X,\ell}(\hat{y}=x-\tilde{z}|x,\ell)\\
& = e^{-\ell \delta \tilde{z}},\quad  0\leq \tilde{z} \leq x, 0\leq \ell \leq K.
\end{align}
Marginalizing over $X$ and $\ell$ leads to:
\begin{align}
CF_{\tilde{Z}}(\tilde{z}) &= \sum_{\ell=0}^{K} \mathbb{P}(K=\ell) \int_{x=\hat{y}}^{\infty} CF_{\tilde{Z}|X,\ell}(\tilde{z}|x,\ell) f(x|\ell) dx \\
&=\sum_{\ell=0}^{K}  \theta^{K-\ell}(1-\theta)^{\ell} \int_{x=\hat{y}}^{\infty} e^{\ell \delta(\hat{y}-x)} \lambda e^{-\lambda x} dx \\ 
&= \sum_{\ell=0}^{K}  \theta^{K-\ell}(1-\theta)^{\ell} e^{ - (\lambda + \ell \delta) \tilde{z}}\\
&= \frac{e^{z(\delta-\lambda)} \theta^{K+1} 
- e^{z(\delta K + \lambda)}(1-\theta)^{K+1}
}{ 1- \theta(1+e^{\delta z})}.
\end{align}

The PDF of $\tilde{Z}$ can now be found by differentiating $1-CF_{\tilde{Z}}(\tilde{z})$ wrt.\ $z$. It is easy to show that this distribution is not exponentially distributed.

\section{Conclusions}
We considered a memoryless one-sided exponential source $X$, which was optimally encoded into $K$ parallel channels $(Y_1,\dotsc, Y_K)$ under the one-sided error criterion, and decoded using the simple select-max estimator, i.e., $\hat{Y} = \max (Y_1, \dotsc, Y_K)$. We showed that the reconstruction error $X - \hat{Y}$ was independent of the outputs $(Y_1, \dotsc, Y_K)$ and was one-sided exponentially distributed. Thus, the statistical properties of the proposed $K$-channel setup can be modelled by an equivalent backward test channel, where the distribution of the reconstruction error is identical to that of the noise in the test channel. This parallels the well-known case of a $K$-channel Gaussian scheme under  MSE distortion and using linear estimation for decoding, which also has an equivalent backward test-channel formulation. 

\bibliographystyle{IEEEtran}
\bibliography{refs}

\end{document}